\documentclass{article}
\usepackage{ptdr-definitions}
\usepackage{hepunits}
\usepackage{heppennames2}
\usepackage{xpatch}
\makeatletter
\xpatchcmd\@HepConStyle
 {\edef\@upcode{\updefault}}
 {\ifdefined\shapedefault\edef\@upcode{\shapedefault}\else\edef\@upcode{\updefault}\fi}
 {}{}
\makeatother
\RequirePackage[colorlinks,citecolor=blue,urlcolor=blue,linkcolor=blue]{hyperref}
\RequirePackage[font=11pt,paper=letter,lang=american]{hep-paper}

\bibliography{bibliography}

\title{
The CMS Heavy Ion Group contribution to\\{\medskip\large
2022 NSAC Long-Range Plan Town Hall Meeting (Hot and Cold QCD) -- Letter of Interest}}

\author[1]{Georgios K. Krintiras and Andr{\'e} G. St{\aa}hl Leit{\'o}n (for the CMS Heavy Ion Group)}

\pdfmapfile{=cid-base.map}
\begin{document}

\maketitle

\noindent {\large \bf Thematic Areas:}

\noindent $\blacksquare$ Physics opportunities at HL-LHC (Runs 3--4) \\
\noindent $\blacksquare$ Upgrade opportunities
 \\
\noindent $\blacksquare$ Synergies with other collider projects  \\
\noindent $\blacksquare$ Physics case at HL-LHC (Run 5 and beyond) \\
\noindent $\blacksquare$ Proposed ``Future Performance'' studies for the White Paper \\

\noindent {\large \bf Contact Information:}\\
CMS Heavy Ion Group Conveners [cms-phys-conveners-HIN@cern.ch]\\

\noindent
The heavy ion (HI) program at the LHC has proven to be a successful and indispensable part of the LHC physics program. Its chief aim had been the detailed characterization of the quark-gluon plasma (QGP) in lead-lead (PbPb) collisions~\cite{CMS:2011iwn}. Using additional data sets of proton-lead (pPb), proton-proton (pp), and xenon-xenon (XeXe) collisions, the program has also included many advances, for example, in the understanding of the partonic nuclear structure~\cite{CMS:2013jlh}, collectivity in smaller collision systems~\cite{CMS:2010ifv,CMS:2012qk},
and electromagnetic (EM) interactions~\cite{CMS:2022arf}. This Letter of Interest outlines the CMS Heavy Ion Group point of view regarding the scientific case for the use of ultrarelativistic HI beams in the coming decade to characterize QGP with unparalleled precision and to probe novel fundamental physics phenomena. More specifically, it outlines the open questions in the field which can be addressed with CMS, and aims to promote engagement from the US community and its international partners by building upon the recently concluded Snowmass 2022 exercise~\cite{CMS:2022cju}, the input provided to the European Strategy for Particle Physics~\cite{Citron:2018lsq}, and proposed continuations and extensions of the last version of the US Long-Range Plan for Nuclear Physics~\cite{Akiba:2015jwa}.

The CMS Collaboration is actively developing the physics program for the High-Luminosity LHC (HL-LHC). The HL-LHC will extend the ongoing LHC program to the second half of the 2030s with pp collisions at a center-of-mass energy of $\sqrt{s}=14\,\TeVns$. At the same time, HL-LHC will collide PbPb and pPb systems at $\sqrtsNN=5.5$ and 8.8\,\TeVns with target integrated luminosities of 13\,\nbinv  and 1.2\,\pbinv, respectively.  The LHC Run 5 could also feature collisions of lighter ions, in the same vein as previously-executed XeXe collisions, which have been predicted to result in substantially increased integrated nucleon-nucleon luminosities. 

In addition to the larger available luminosity, detector upgrades planned for the CMS experiment will benefit the HI program. In particular, an increased charged particle
tracking pseudorapidity ($\eta$) acceptance resulting from tracker upgrades will be a boon to bulk particle measurements, and the upgraded Zero Degree Calorimeters (ZDC) will improve triggering and identification of ultraperipheral
collisions (UPC). The addition of time-of-flight particle identification capability enabled by the Minimum Ionizing Particle (MIP) Timing Detector~\cite{CMS:2667167} will allow for differentiation among low momentum charged hadrons, such as pions, kaons, and protons, which will improve measurements of heavy flavor particles and neutral strange hadrons such as $K^0_s$ and $\Lambda$~\cite{CMS-DP-2021-037}.

The predicted factor of ten increase in the integrated luminosity compared to the initial LHC program, combined with new and improved detector components, will open the door for refined studies of QGP properties and complement the planned trajectory of the LHC pp program. This underscores the wealth of physics awaiting study at the LHC and its detectors, which will extend far beyond original design goals.
Fully exploiting these exciting opportunities requires synergies among experts in the accelerator, experimental, and theory communities, as well as long-term commitments to the development of relevant technologies for the lifetime of the LHC program.

We deem the following physics topics particularly well suited to be covered, in a unique way, by general purpose detectors like the CMS experiment:

\paragraph{Physics opportunities at HL-LHC (Runs 3--4)}
\subparagraph{Initial-state physics}

The structure of bound nucleons in nuclei is key to understanding fundamental nuclear modifications as well as the initial state of HI collisions. Measurements made during the LHC Runs 1 and 2 in PbPb and pPb collisions~\cite{Amoroso:2022eow}, have allowed for the inclusion of nuclear modifications to the parton distribution functions (PDFs) extracted from free protons. In particular, the pPb collision system is a valuable tool to constrain the parameter space in the formulated nuclear parton distribution functions (nPDFs) since its asymmetric nature allows for probing poorly-constrained low Bjorken-$x$ values by looking at the proton-going direction.  

To this end, CMS will measure with unprecedented precision \PW and \PZ boson production.  Differential measurements in rapidity (or lepton $\eta$ in the \PW boson case) using pPb collisions will constrain the quark (valence or sea) and gluon nPDFs~\cite{FTR-18-027}. Complementing these measurements, CMS projects the ability to make very precise measurements of dijet pseudorapidity, which is an observable highly sensitive to the gluon nPDF~\cite{FTR-18-027}. Furthermore, the measurement of differential \ttbar cross sections in pPb collisions is a novel and potentially precise probe of the nuclear gluon density~\cite{FTR-18-027}.  Such a measurement requires a great deal of delivered luminosity to execute, meaning large improvements are expected during the course of the HL-LHC era.  Finally, by taking advantage of heavy quark production associated with vector bosons in pPb collisions (similar to, \eg, previously studied $\PW{}+\PQc$ and $\PW{}+\PQb$ processes in pp collisions), global nPDF fits could be improved with sensitivity to the \PQs and \PQc quark nPDFs, respectively. 

In the very small-$x$ region, the resummation of 
$\ln(1/x)$ contributions can be handled in the BFKL framework. 
Additionally, parton saturation and recombination are 
expected to grow in this region with a $A^{1/3}$ enhancement
for heavy nuclei. If the gluon saturation regime is reached, as expected, at some value of $x$, then collinear factorization breaks down, and the (n)PDF concept breaks down. The latter is strongly related to the formation of an over-occupied system of gluons, \ie, the Color Glass Condensate, whose exploration is one of the central physics goals of a future Electron Ion Collider (EIC). At an EIC a dense QCD state will be probed through the scattering of electrons on heavy ions.  However, forward physics at LHC can also explore this dense QCD state since high gluon densities are produced through the low-$x$ evolution of the gluon distribution of the Pb ion~\cite{Hentschinski:2022xnd}.  In this way, the forward physics program at the LHC is expected to be complementary to the physics program at an EIC.

\subparagraph{\textbf{Particle collectivity in small and large systems}}
By leveraging upgrades to experimental data-acquisition systems during the HL-LHC era, the CMS experiment plans to gather unprecedentedly large samples of minimum bias pp, pPb, and PbPb collisions.  With these upgrades it will be possible to improve experimental precision over an extended final-state particle multiplicity range, which will help us understand the origin of collectivity in nuclear collisions, as well as smaller systems such as pp.  This will also help constrain the transport properties of the QGP.  

In small systems, this will result in a significant statistical improvement for more elaborate flow variables that are sensitive to the initial state and its subnucleon fluctuations. For example, the so-called subevent method, which significantly suppresses nonflow contributions, will help us examine collectivity down to very low multiplicity and study the onset of flow in small systems~\cite{FTR-18-026}.  Using azimuthally sensitive femtoscopy techniques, improvements are expected in our understanding of the system size of collisions by measuring the Hanbury Brown and Twiss radii in small systems. 

In PbPb collisions, higher order flow harmonics, as well as observables that quantify correlations between different harmonics, show larger sensitivity to the medium properties. All those observables are statistically demanding and will be highly enriched by HL-LHC data. The new data will also open the door for event-by-event correlation measurements that include heavy flavor particles and jets into the already-established procedures for measuring bulk-produced particles. Such correlations will clearly reveal energy loss mechanisms inside the QGP.  Furthermore, similar measurements of femtoscopic correlations between strange hadrons and nucleons could aid understanding of the composition of neutron star cores.

Flow measurements will not benefit only from higher statistics during the HL-LHC era.  Upgrades of the CMS tracker will also enable the measurement of charged particles in a wider pseudorapidity range ($\abs{\eta}<4$).  The extended $\eta$ acceptance will lead to significant improvement in characterizing the rapidity dependence of correlations, which will allow flow decorrelation~\cite{Gardim:2012im} and forward-backward multiplicity correlation studies over a wide rapidity range.  Such measurements will help set constraints on the fluctuations of the medium in the early stages of a collision.

\subparagraph{Jets, high-\pt hadrons, and QCD medium response}

Studies of jet modification in HI collisions provide information on the interaction between high energy partons and the QGP. The large data samples and the improved jet reconstruction resulting from CMS tracker and calorimeter upgrades will provide significantly reduced statistical and systematic uncertainties for key measurements of medium modification of light (heavy) quark jets using \PGg/\PZ (\PDz meson) tagged samples~\cite{FTR-18-025}. 
By measuring the jets tagged against recoiling isolated photons, one can unambiguously access the energy loss of a parton by using the photon/\PZ energy as the reference for that of the parton before quenching.  In particular, the jet fragmentation functions and the jet shapes will be measured precisely where the difference between pp and PbPb collisions is not resolved yet~\cite{FTR-18-025}, hence providing information on the medium-modified structure of quark-initiated jets. The azimuthal and \pt correlations between the electroweak bosons and jets, measured in \PGg/\PZ+jets events, are valuable observables to study parton energy loss in the QGP~\cite{FTR-18-025} which can be probed using HL-LHC data. Another significant development expected at the HL-LHC is the measurement of radial distribution of \PDz mesons in jets~\cite{FTR-18-025}. By studying the modification of this observable in PbPb compared to pp, one can gain insights into the dynamics of heavy quarks in the QGP. 

Present measurements at the LHC, \eg, Ref.~\cite{CMS:2021vui} have shown the jet structure to be modified for a series of typical jet cone sizes ($R$ = 0.1--1), with the fragmentation products inside the cone shifted towards lower \pt and larger angles, and associated transport of most of the quenched jet energy outside of the jet cone size. Meanwhile, there is an intense ongoing activity in developing generalized jet (sub)structure variables at the LHC to maximize the efficiency of discovery measurements by improving quark/gluon discrimination and the tagging of boosted objects. We anticipate that heavy ion studies of the modification of the jet momentum and angular structure through medium interactions will benefit greatly from these developments, especially when combined with high-precision data from the HL-LHC.

\subparagraph{Heavy flavor, quarkonia, and strangeness production }
Heavy flavor (HF) hadrons are excellent probes of the strongly interacting medium. Charm and beauty quarks, as a consequence of their large masses, are mainly produced during the early stages of the collision. In addition, quarkonia can be dissociated or recombined in the medium. Therefore, the measurement of HF hadrons can provide crucial information on the full evolution of the system and  the quark-mass dependence of the medium-induced processes. The larger experimental data samples at the HL-LHC, combined with improved detector performance and measurement techniques, will allow for significant improvements over the current HF measurements. The \pt dependence of the quarkonium nuclear modification factor ($R_{\mathrm{AA}}$) will be measured with high precision up to about 80\,\GeVns for prompt \PJGy and 50\,\GeVns for $\Upsilon(1S)$~\cite{FTR-18-024} (compared to 50 and 30\,\GeVns with the present data, respectively).  The centrality dependence of $\Upsilon$ states, including the recently observed $\Upsilon(3S)$~\cite{CMS:2022rna}, will be measured with much greater accuracy.  Similar improvements in kinematic reach are also expected for open charm (\eg, $\PD^{0}$) and beauty mesons (e.g. $\PB^{+}$), allowing for stringent tests of models containing different energy-loss mechanisms for heavy flavor quarks.  The production of strange \PB mesons and charm baryons (\eg, $\Lambda_c^+$) in pp and PbPb collisions~\cite{FTR-18-024} will also be measured with sufficient precision to further investigate the interplay between the predicted enhancement of strange quark production and the quenching of beauty quarks, as well as the contribution of recombination of HF quarks with lighter quarks to the hadronization process in HI collisions.   Elliptic flow measurements of  $\Upsilon(1S)$ mesons~\cite{FTR-18-024} in PbPb collisions and charm mesons in pPb collisions~\cite{FTR-18-026} will be significantly improved.   Finally, precise measurements of beauty mesons in pPb collisions~\cite{FTR-18-024} will help to elucidate the relative contribution of hadronization and nuclear-matter effects (\eg, nuclear PDF, gluon saturation, and coherent energy loss), as well as serve as a baseline for the understanding of \PQb quark energy loss in PbPb collisions.

Particles with an alternative quark content, known as ``exotic'' states, have been actively discussed since the birth of the QCD parton model, with the discussion revived by a series of more recent observations of numerous tetraquark and pentaquark candidates (see Refs.~\cite{CMS:2021znk,LHCb:2021vvq} for a review). Due to the closeness of their masses to known particle-pair thresholds, many of these states are likely to be hadronic molecules, albeit an interpretation of exotic states as compact multiquark structures is also possible.  Particles that are loosely bound hadronic molecules are expected to more easily dissociate in the QGP as compared to compact multiquark states.  Thus, future LHC Run 3 and 4 data sets enable a unique opportunity to provide additional constraints regarding the nature of these exotic states.

\subparagraph{Ultraperipheral collisions}
In addition to hadronic collisions, which have been the main focus of the HI program at the LHC, the large electromagnetic fields generated by ultrarelativistic charged ions can interact with the nucleus (photo-nuclear processes) or with each other (photon-photon interactions). Photo-nuclear collisions probe the nuclear structure and can constrain nPDFs. On the other hand, photon-photon interactions can be used to calibrate high-uncertainty inputs to QED calculations (\eg, the so-called photon flux) and to the search for physics beyond the standard model~\cite{dEnterria:2022sut}. The CMS Collaboration has a suite of planned measurements~\cite{FTR-18-027} (photoproduction of vector mesons~\cite{CMS:2018bbk,CMS:2019awk}, exclusive lepton pair production~\cite{CMS:2020skx}, and light-by-light scattering process~\cite{Sirunyan:2018fhl}) on these topics.  These measurements will all benefit from planned ZDC upgrades for better triggering capabilities, segmentation, and radiation hardness.

\paragraph{CMS Phase 2 Upgrade opportunities}
\subparagraph{The MIP Timing Detector for HIC}
The large variety of physics topics examined during the LHC Runs 1 and 2, as well as the proposed program outlined above for the HL-LHC, illustrates the remarkable versatility and strengths of the CMS detector. To complement these existing advantages, a new timing detector, MIP timing detector (MTD), will be in operation during HL-LHC from Run 4~\cite{CMS:2667167}. It will equip the CMS detector with a high timing precision in a unique pseudorapidity range of up to six units. The designed precision is expected to be 30 ps during Run 4, and is expected to degrade to around 50 ps by the end of HL-LHC era.

The mechanical structure of the MTD consists of barrel timing layers (BTL) and endcap timing layers (ETL). 
BTL is a cylindrical detector, occupies the inner surface of the Tracker Support Tube, and has a thickness of 40 mm. The inner radius of BTL is 1148 mm and longitudinal length spans 2.6 meters on each side with respect to the center of CMS detector. It covers the pseudo-rapidity range $|\eta|<1.45$.
ETL is located on the endcap calorimetry nose where the longitudinal displacement on each side is 3 meters with respect to the detector center and the radius spans from 315 mm to 1200 mm. ETL fits into the small gap of 45 mm between tracker and endcap calorimetry. It covers the pseudo-rapidity range $1.6<|\eta|<3.0$.

In addition to differences in geometry, BTL and ETL are developed with different detector technologies. For the barrel part, crystal scintillator LYSO:Ce bars are used and the readout system employs the technology of silicon photon multiplier (SiPM). Its surface area is 38 $\mathrm{m}^2$ with 331776 readout channels. The radiation tolerance can be up to “1 MeV neutron equivalent fluence” of $2 \times 10^{14} \mathrm{n_{eq}}/\mathrm{cm}^{-2}$. On the other hand, ETL is constructed using Low Gain Avalanche Detectors (LGADs) because of severe radiation conditions. ETL can tolerate the radiation up to $2 \times 10^{15}\mathrm{n_{eq}/cm}^2$. The LGADs have an internal gain of 10-30 provided by special implants. Each disk of the endcap is occupied by modules constructed by 16x32 arrays of square LGAD pads of $1.3 \times 1.3 \mathrm{mm}^2$. The active area of ETL is 14 m$^2$ with about 8.5 million channels. 

The time-of-flight measured by MTD provides the opportunity to identify hadrons in heavy ion collisions. It will allow measurements of the flow of identified particles through long-range two- and multi-particle correlations and will shed light on the initial stage of a collision and the inner workings of collectivity in large and small systems, as discussed in the previous section. Other advantages include the enabling of studies of inside- and outside-of jet particle chemistry and dynamics through measurements of baryon and meson yields. The MTD also benefits heavy flavor particle reconstructions by suppressing combinatorial backgrounds through particle identification. For instance, CMS will be able to access the total charm cross section in a large phase space and measure the collectivity of charm hadrons precisely.

Thus, when combined with other CMS Phase II detector upgrades, the MTD provides unprecedented opportunities for heavy ion physics during HL-LHC era.

\paragraph{Synergies with other collider projects}

\subparagraph{Cold and Hot QCD at RHIC and LHC}
The combination of large increases in delivered luminosity over the next decade, upgrades to the existing LHC detectors, and the operation of a state-of-the-art jet detector at RHIC (\ie, sPHENIX) will enable a coherent physics program employing well-calibrated common observables to study jet
modifications and jet-medium interactions over a wide range of medium conditions. The different energy regimes and tagging of particular initial states (\eg, \PGg+jet, \Pb-tagged jets, dijet
events), precision measurements of modifications of the jet (sub)structure in angular and momentum space, potential modifications to the back-to-back jet scattering distributions, as well as modifications of the intrajet angular structure will allow the selection of jet populations in relation to different medium conditions. As jets and the medium co-evolve from their initial virtuality and transform to final-state hadrons, the comparison of various observables between the RHIC and LHC events will provide key insights into the temperature dependence of the jet-medium interactions. Success in this long-term endeavor will require a global analysis of a
diverse set of RHIC and LHC data in an improved, well-controlled theoretical framework that makes an explicit interface with the experimental observables.

The different color screening environments caused by the different temperatures attained in collisions at RHIC and LHC, combined with the large differences in the time evolution of the QGP
and in the underlying bottom production rates, make them also distinctly different laboratories for
studying the effect of the QGP on the $\Upsilon(\mathrm{nS})$ states. The combination of high precision $\Upsilon(\mathrm{nS})$ data
from the LHC and RHIC will constrain theoretical models in a way that data measured at only one energy could otherwise not.

\subparagraph{Cold QCD at EIC and LHC}
The construction of an Electron-Ion Collider (EIC)~\cite{AbdulKhalek:2021gbh} has been approved by the United States
Department of Energy at Brookhaven National Laboratory, and could record the first scattering events from (polarized) electron, and potentially positron, beams with proton or ion beams at a center-of-mass
energy of up to $\lesssim 140$\,\GeVns as early as 2030. There is a significant cross-talk between the research areas at the Intensity (EIC) and Energy (LHC) frontiers in our effort to understand how the nuclear medium modifies parton-level interactions and substructure, whether gluons saturate within heavy nuclei, and in a range of (n)PDF-related precision measurements in fundamental QCD and electroweak phenomenology. Similar to the proposed physics case in LHC Run 5 (cf. below), the EIC will have the capability to operate with a range of nuclei, from deuterium to lead. Thus, the dependence of nPDFs on the atomic mass number could be investigated in a complementary way by both facilities. The abundance of HL-LHC and EIC measurements will make it possible to determine nPDFs independently for each nucleus. Furthermore, the next generation of nDPFs will be improved because, at the EIC, proton and ion beams will be used in
a consistent experimental framework allowing for a combined, simultaneous determination of proton and nPDFs.  

It should also be noted that the LHC hot QCD program will benefit heavily from the expected leap in understanding of the internal structure of heavy ions, and therefore of the initial state of heavy ion collisions, resulting from EIC measurements.  Thus, new opportunities and in some cases significant reductions of systematic uncertainties for LHC measurements can be expected as a result of interaction between the hot and cold QCD communities during the EIC era.

\paragraph{Physics case for HL-LHC (Run 5 and beyond)}
Similar to the Beam Energy Scan operations at RHIC~\cite{Liu:2022wme}, a unique opportunity to study the emergence of collective phenomena as well as parton energy loss in small collision systems is offered by lighter nuclei at LHC. The observation of energy loss in small systems would unambiguously confirm the role of significant final-state interactions as underlying mechanism. On the contrary, the absence of suppression despite possible collective features cannot be taken as a proof against the QGP formation, \eg, partons might be emitted close to the QGP surface. Lighter ions, with the possibility to achieve relatively large integrated luminosities in modest running times (although the exact amount is a topic of active research) offer additional initial-state opportunities, including studies of small-$x$ physics and the precise determination of nPDFs specially due to the scarcity of nuclear data at high $Q^2$.

Since lighter ions potentially offer relatively large luminosities they are also important for heavy flavor production, in particular beauty mesons and bottomonia, and observables with large scales, like electroweak bosons, dijets, and top quarks.  The larger nucleon-nucleon
luminosity could also possibly benefit studies in UPC events despite the suppression of photon-photon luminosities.

Last but not least, large low-pileup pp data samples at $\sqrt{s}=$14\,\TeVns will be a great opportunity for precision measurements of the system-size dependence of the jet quenching phenomena. In this regard, high-multiplicity pp events provide the reference results for small systems, and are therefore important for examining the system-size dependence of the jet quenching phenomena.

\paragraph{Proposed ``Future Performance'' studies for the White Paper}

The CMS physics contribution to the Long Range Plan process is expected to build upon the studies that are summarized in the Snowmass 2022~\cite{CMS:2022cju} community exercise, the CERN Yellow Report on the Physics at the HL-LHC, and Perspectives for the HE-LHC~\cite{Citron:2018lsq}. Below is a nonexhaustive list of the proposed contributions, separated by the identified main physics topics of:
\begin{itemize}
    \item Understanding the low-$x$ gluon density, universarilty of saturation and new insights on the gluon distribution, and magnetohydrodynamical phenomena (\eg, strong magnetic fields in PbPb, QGP electric conductivity) of QCD
    \begin{itemize}
    \item
    Low- and high-mass vector meson production in UPC events
    \item Dimuon and Breit-Wheeler processes in UPC events
    \item Dijet production in UPC (photonuclear or photon-proton interactions)
    \item \PDz meson $\upsilon_1$
    \end{itemize}
    \item Understanding (un)identified particle production mechanisms (\eg, modification of strangeness production, late hadronic rescattering), event-by-event mean \pt fluctuations, flow factorization, and nonflow effects
    \begin{itemize}
    \item Higher order cumulants in ultracentral collisions 
    \item Two particle angular correlations with multi-strange, charm, and bottom hadrons
    \item Femtoscopic correlations and charge balance functions for extensive list of particles (\eg, \PGf meson) and higher mass (\eg, \PGO) baryons 
    \item $\mathrm{\Delta}\pt\mathrm{\Delta}\pt(\Delta\eta\Delta\phi$) correlations
    \end{itemize}
        \item Understanding hadronization, jet quenching, and in medium modifications 
    \begin{itemize}
    \item $x_j$-binned jet substructure in $\gamma$-jet events
    \item Simultaneous measurement of jet $R_\mathrm{AA}$ dependence at inclusive, $\gamma$-jet, and \PQb jets
    \item Electroweak boson tagged jet hadron correlations
    \item Top quark production at high \pt
    \end{itemize}
    \item
   Understanding heavy flavor particle production, correlation dynamics, and energy loss mechanisms
    \begin{itemize}
    \item \PDz and \PDps $\upsilon_n$, $R_\mathrm{AA}$, and \PDps/\PDz meson ratio
    \item Two-particle heavy flavor correlations (\eg, \PDz-\PDz, \PDps-\PDps, \PDps-\PDz)
    \end{itemize}
    \item Understanding exotic hadron composition targeting to a direct comparison with Lattice QCD calculations and even the composition of the core of neutron stars 
    \begin{itemize}
    \item $\mathrm{X}(3872)$ $R_\mathrm{AA}$ in centrality bins, also its photoproduction
    \item $\mathrm{T}_{\PQc\PQc}^+$ production
    \item Femtoscopic correlations between strange hadrons and nucleons 
    \end{itemize}
    \item Understanding the physics of ultraperipheral collisions with a view beyond the standard model
     \begin{itemize}
    \item
    \PGt lepton $(g-2)$ with an absolute uncertainty of less than 0.01
    \item Light-by-light scattering and searches for low mass ($\lesssim100$\,\GeVns) axion-like particles
     \end{itemize}
  \end{itemize}

\paragraph{Outlook} 

The QCD theory of strong interactions remains one of the critical components of the standard model to be properly understood.  The large \alpS value at low $Q^2$ renders traditional small-\alpS perturbation theory inapplicable, such that collective phenomena in nuclei are nonperturbative and therefore not easily handled by a computational framework. However, a coordinated application of the QCD parton model for conventional hadrons, an effort to grasp the exotic hadron spectroscopy, and advances from Lattice QCD calculations could yield a fundamentally improved understanding of the characteristics of nuclei and their interactions.

At the same time, much remains to be learned about the precise nature of the initial state from which the thermal QCD matter forms. Questions remain about parton densities in nuclei in a broad $(x, Q^2)$ kinematic range, the search for the possible onset of parton saturation and how its properties vary across its phase diagram, and how the collective properties of this quark-gluon plasma emerge at a microscopic level from the interactions among the individual quarks and gluons that must be visible if the liquid is probed with sufficiently high resolution. Therefore a crucial aspect of nuclear studies is the exploitation of future opportunities for high-density QCD studies with ion and proton beams.  This will allow the study of cold nuclear matter effects, the onset of nuclear saturation, and long-range correlations.  Additionally, examination of high-\pt hadrons, fully reconstructed jets, heavy quarkonia, open heavy flavor particles, and jet quenching will provide information about the strongly coupled QGP, complementing the bulk and collective observables of the soft sector.  It is also important to understand at what level these various phenomena might be phenomenologically limited by present and future knowledge of the nPDFs.

As illustrated by the CMS heavy ion community's future plans, the continuation of the LHC heavy ion physics program into the HL-LHC era is a unique opportunity which should be supported in the upcoming decade.  This program has strong complementarity with other key research efforts in the nuclear physics QCD community (i.e., existing efforts at RHIC and the upcoming EIC), synergizes with technical developments in the high-energy physics community, and will play a key role in furthering the goal of strengthening our understanding of both QED and QCD.

\printbibliography
\end{document}